\documentclass[aps,pra,showpacs,twoside,twocolumn,10pt]{revtex4-1}
\usepackage[colorlinks=true, citecolor=red, urlcolor=blue ]{hyperref}
\usepackage{xcolor}
\usepackage{epsfig,newlfont,amssymb,amsfonts,amsmath,bm,subfigure,palatino,mathtools,amsthm,braket,times,soul,enumitem,color}
\usepackage[normalem]{ulem}
\newcommand{\stkout}[1]{\ifmmode\text{\sout{\ensuremath{#1}}}\else\sout{#1}\fi}

\usepackage[utf8]{inputenc}
\usepackage{array}
\usepackage{graphics}

\newcommand{\ketbra}[2]{|#1\rangle \langle #2|}
\begin{document}

\title{
Restoring metrological quantum advantage of measurement precision in noisy scenario 
}
\author{Aparajita Bhattacharyya, Ahana Ghoshal, Ujjwal Sen}
\affiliation{
Harish-Chandra Research Institute, A CI of Homi Bhabha National Institute, Chhatnag Road, Jhunsi, Prayagraj 211 019, India}

\begin{abstract}
We show that in presence of a local and uncorrelated dephasing noise, quantum advantage can be obtained in the Fisher information-based lower bound of the minimum uncertainty in estimating parameters of the system Hamiltonian. The quantum advantage refers here to the benefit of initiating with a maximally entangled state instead of a product one. This quantum advantage was known to vanish in the same noisy scenario for a frequency estimation protocol.
Restoration of the better precision in frequency estimation with maximally entangled probes can be obtained by incorporating an interaction between the system particles. The interaction examined here is Ising in nature, and is considered with or without a transverse magnetic field. There are instances, e.g. where frequency estimation in presence of a transverse field is considered and quantum advantage is not restored.
A quantum advantage can also be obtained while estimating the strength of the introduced magnetic field along the transverse direction, whereas for the instances considered,  using uncorrelated probes is better in measuring the coupling parameter of the Ising interaction. We also investigate the dependence of measurement precision on the entanglement content, which is not necessarily maximal, of the initial state. The precision in estimation of coupling constant decreases monotonically with the increase of entanglement content of the initial state, while the same for frequency estimation is independent of the entanglement content of the inputs.
\end{abstract}

\maketitle
\section{Introduction}
Quantum metrology, which deals with the enhancement - with respect to ``classical" means - in the sensitivity of measurement of a physical quantity~\cite{Braunstein1,Lloyd0,Treutlein}, 
has been extensively studied 
in the last two decades and is facilitated by quantum resources~\cite{Cirac&Plenio,Maccone,Giovannetti_Maccone,Pirandola} like  entanglement~\cite{Plenio1,Horodecki,Toth,Lewenstein1}.
The developments in the field have far-reaching consequences in various arenas of physics like gravitational
wave detection~\cite{astro}, optical imaging under high resolution~\cite{q_imaging,Genoni3,Liang}, quantum thermometries~\cite{therm1,therm2}, magnetometers~\cite{mag1,mag2}, etc.
 Entanglement is widely used as a resource to improve the metrology precision~\cite{Cirac&Plenio,Maccone,Lloyd0}. It has been observed that quantum entanglement can be utilized to overcome the so-called shot-noise limit 
~\cite{Cirac&Plenio,Lukin,Nagata,Roy,Lewenstein,Matsuzaki,Hu} obtained without quantum resources, often referred to as the ``classical limit".
The classical limit sets the lower bound of the uncertainty of measuring a physical observable in an experiment, without quantum resources, attainable in the asymptotic limit, and proportional to the inverse square root of the number of measurements.

In any measurement process, the origin of the errors can be of two different types. One of them arises fundamentally from the Heisenberg uncertainty principle, while the other is due to lack of control over the system or the probes. These errors can be minimized using certain quantumnesses incorporated into the system in the form of entanglement or squeezing~\cite{Giovannetti_Maccone2}.
For example, it has been shown that the classical bound - the shot-noise limit - can be overcome by utilizing the quantum nature of entangled photons~\cite{opt1,opt2,opt3,opt4,opt5,opt6}. The relvant measurements can be executed using different interferometers~\cite{Holland,opt7} like the Ramsey spectroscope, the Mach-Zehnder interferometer~\cite{Dowling,Sanders,Ou}, etc. Particularly, in some interferometric experiments which detect the changes in state population, the signal-to-noise ratio can be improved by using quantum effects using spin squeezed states~\cite{Heinzen}. The frequency estimation has been generalized in~\cite{Shaji}, where probe generation rate and evolution time were considered as resources.

A certain lower bound in evaluating the variance of a parameter classically, commonly known as the Cram\'er-Rao bound, was obtained by using a quantity, called the Fisher information, which provides a way of measuring the amount of information that a random variable contains about an unknown parameter,  on which the probability distribution of the random variable depends~\cite{Cramer}. 
The quantum version of the Cram\'er-Rao bound can be obtained by utilizing the quantum Fisher information (QFI)~\cite{Helstrom,Braunstein,Milburn,Holevo,Hayashi1,Hayashi2,Paris_book,Paris4}, where a maximization  over all possible measurements is involved. 


In presence of decoherence during time evolution, the metrological performance of a system generally deteriorates and is considered to be one of the main hurdles in entanglement-enhanced sensing~\cite{Cirac&Plenio,Saunders}, but sometimes decoherence shows some benefits. It was shown that under suitable conditions, a decoherence effect, specifically the effect of a Markvoian collective dephasing channel~\cite{Matsuzaki2}, can be utilized to enhance the sensitivity of measurement.
Recently, non-Markovian effects have also been used to attain high-precision in quantum
optical metrology under locally dissipative environments~\cite{An}.
It is known that the metrological error gradually surges up in the
long-encoding-time regime under the influence of decoherence, caused by environments~\cite{Paris3,Kurizki,Genoni,Huelga,Paris,Paris2,Wang,Binder,Agarwal}. 
This circumstance is called the no-go theorem of noisy quantum metrology~\cite{Genoni,Wang,Binder,Agarwal,Huelga2} and is a major barrier in attaining high-precision quantum metrology in practice. However, 
it has
been revealed that a non-Markovian calculation can obtain a qualitatively distinct dynamical behavior and it helps to surpass the no-go theorem~\cite{Huelga3,Berrada,Shi,Tong_Luo,Matsuzaki_new,Xu,Lin}. Estimation of channel parameters of a noisy quantum channel~\cite{Fischer} is another domain of work in metrology. See also~\cite{Gill} for state estimation and~\cite{Fujiwara} for channel identification problems.

Several strategies have been introduced to  
exceed the shot-noise limit and achieve better precision.
These include the use of non-linearity in the system~\cite{Boixo,Boixo1,Boixo2,Choi,Chase,Jacob,Tilma,Mitchell}, squeezing of the vacuum ~\cite{Caves,Dowling2,Schnabel,Schnabel2,Adhikari}, optimization of the probing time~\cite{Chaves}, controlling the environment~\cite{Plenio&Huelga2,Genoni2} and non-Markovian evolutions~\cite{Huelga2,Huelga3}.
However, for every quantum resource, there is a non-trivial bound to the corresponding quantum advantage. 
A bound can also be derived in postselected metrology~\cite{Lloyd}, setting up an inter-relation
with weak value optimization, where the latter can be realised in terms of a geometric phase~\cite{Bera}.
The scaling in the variance of the estimated parameter in terms of the quantum Fisher information in the asymptotic limit, in  many-body open quantum systems has also been studied~\cite{Campo,kbody}.

An important point to be noted here is that the interaction incorporated into the Hamiltonian of the system can have a crucial impact on the precision limit, as demonstrated in Ref.~\cite{Boixo2}. 
They have shown that the metrological precision scaling that can be obtained by using a probe, described by the Hamiltonian $H=\mathcal{J}_z^2$, with $\mathcal{J}_z=\sum_{i=1}^N\sigma_i^z$ is the same as that can be achieved by doing it with $H=N\mathcal{J}_z$. Here, $N$ represents the number of qubits in each probe.
This provides crucial insight about the response of interaction terms on metrological precision. A difference between their approach and ours is that unlike in their approach, the interactions here can be continuously reduced to the case of no interactions. Also, we have analyzed the response of transverse fields on the precision.


\begin{figure}
\centering
\includegraphics[width=8cm]{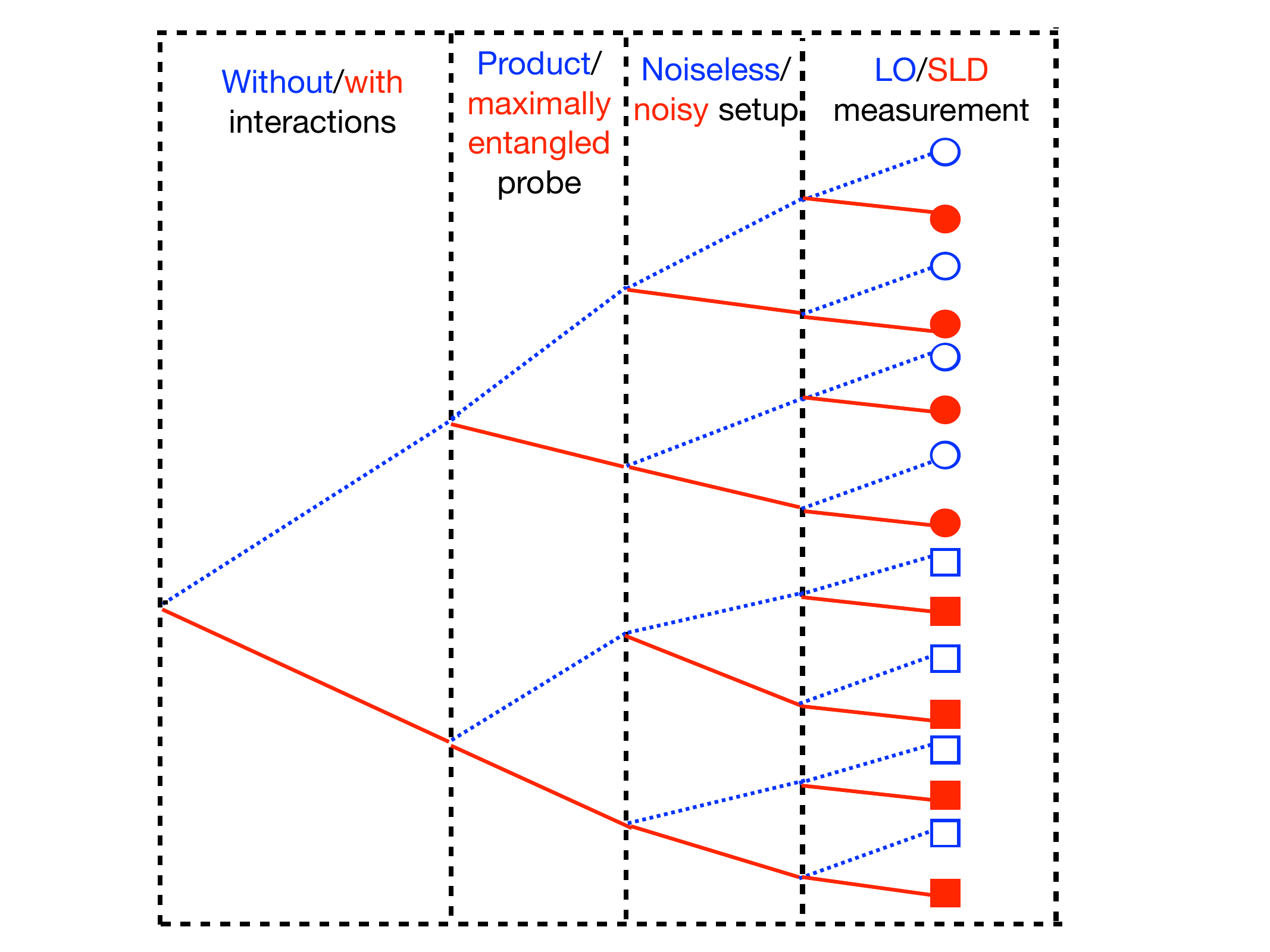}%
\caption{Schematic to depict the 
different situations considered in estimating the minimum deviations of the relevant parameters. So, in particular, the third dot from the top in the rightmost column refers to the case where the probes are encoded via a non-interacting Hamiltonian, the probe is initially in a product state, the setup is noisy, and the measurement is of the LO variety. The circles correspond to the situations where the encoding mechanism is via a non-interacting Hamiltonian. The squares on the other hand correspond to those for which the encoding Hamiltonian is interacting.
In every bifurcation, the upper line is blue dotted and the lower one is red solid. Correspondingly, in the option at the top of every column, the first one is blue and the second is red.}
\label{schemes}
\end{figure}

In this work, we consider probes of two parties throughout the analysis, and we focus on two measurement schemes for the parameter estimation, viz. the strategy in which the measurements are locally carried out at the two parties, and the optimal measurement strategy. The former involves measuring onto a biorthogonal product basis while the latter is in general a measurement onto a more general basis, possibly involving entangled states, and involves consideration of a symmetric logarithmic derivative (SLD). 
We will refer to these strategies as LO and SLD strategies, or as Schemes I and II respectively, for brevity. We will analyze estimation of frequency, or coupling strength of an interaction between the probe parties, or a transverse field on the parties. We consider both noiseless and noisy scenarios, with the noise being of dephasing type. For a depiction of the tree corresponding to the different cases considered, refer to Fig.~\ref{schemes}.

Let us first try to be precise about what is meant in this paper by ``quantum advantage''.
If there is an enhancement in the precision of estimation of a certain parameter using maximally entangled initial probe states instead of unentangled ones, keeping all other aspects of the setup as unchanged, we refer to it as quantum advantage.

For LO measurements, metrological advantage in frequency estimation 
increases by using maximally entangled initial states instead of unentangled ones in the noiseless scenario, but the effect of decoherence hinders the performance of measurement by decreasing its sensitivity. See \cite{Cirac&Plenio} in this regard.

\emph{LO measurements, with interaction.} Let us first consider the noisy case. For frequency estimation, quantum advantage is regained in the noisy case by incorporating an Ising coupling in the encoding Hamiltonian with or without a transverse field, for LO measurements. For estimating the coupling strength, however, using a product probe is better. Estimation of the transverse field provides quantum advantage again, in presence of the interaction between the two parties of the probe. In the noiseless situation,
a maximally entangled probe provides a better precision for frequency estimation in presence of interaction, with or without a transverse field, in the encoding Hamiltonian. The precision of estimating the coupling strength in the noiseless case is immune to whether  frequency and transverse field terms are present in the encoding Hamiltonian. Quantum advantage is however regained in the noiseless setup for transverse field estimation in presence of interaction.

\emph{LO measurements, without interaction.} As above, we begin with the noisy case. There is no quantum advantage in the noisy case for frequency estimation, irrespective of whether we use a transverse field, in absence of an interaction term, when LO measurements are utilized. However, quantum advantage is regained for transverse field estimation. 
In the noiseless case, frequency as well as transverse field estimation possess quantum advantage.

Before proceeding further, let us note here that for frequency estimation in the noisy case using the best measurement strategy (i.e., the SLD strategy), quantum advantage is already present even without any interaction between the parties of the probe or any transverse field.

\emph{SLD measurements, with interaction.} In this case, quantum advantage is present in the noisy situation for frequency estimation, in presence of an Ising coupling, for SLD strategies. This remains so even for an additional transverse field in the encoding Hamiltonian. For estimating the coupling in the same scenario, product and maximally entangled probes lead to identical precision. Quantum advantage is regained however for estimating the transverse field. 
The results are similar in nature for the noiseless scenario.

\emph{SLD measurements, without interaction.} 
In the noisy 
case, quantum advantage prevails both in presence and absence of the transverse magnetic field, while estimating the frequency, and also in the estimation of interaction strength.  Similar are the behaviors for the noiseless situations.

Furthermore, we observe that the measurement precision monotonically decreases with the increase of entanglement content of the initial state for certain field strength estimation protocols, whereas for frequency estimation, the
measurement precision 
is independent of the entanglement content of the initial state. 

The remainder of the paper is arranged as follows. The relevant information from previous literature is discussed in Sec.~\ref{Sec:2}. This includes the methods and the results of measuring the minimum uncertainty of estimating system parameters, which are to be measured in presence and absence of noise, both for the product and maximally entangled initial states. 
Section~\ref{LO} constitutes the significant part of our paper, where we present the results of the minimum uncertainties obtained while measuring  different system parameters in presence of fields and two-qubit interactions, for probes that are product or maximally entangled states initially. In Sec.~\ref{sld0}, we draw a comparison between the results obtained in Sec.~\ref{LO} and the optimal measurement protocol of quantum metrology. In Sec.~\ref{arb_in}, we investigate the dependence of the minimum uncertainty of the system parameters on the entanglement content of the initial state. Concluding remarks are presented in Sec.~\ref{Sec:5}.
\section{Quantum metrology and frequency estimation protocol}
\label{Sec:2}
Metrology pertains to the estimation of unknown physical parameters of the system~\cite{Braunstein1,Holland,Lloyd0}. 
Whenever a procedure using quantum resources outperforms a similar ``classical" process, i.e. the one without the quantum resource, it is said that a quantum advantage has been achieved. The term quantum metrology is used when the estimations are improved in presence of quantum resources~\cite{Maccone,Giovannetti_Maccone,Pirandola}  like entanglement. There are various methods of  estimating a parameter accurately in the asymptotic limit using the concept of quantum metrology. One of them is based on the Cram\'{e}r-Rao bound as proposed in~\cite{Maccone}.

We begin with a situation where we want to estimate a system parameter $\theta$, which is encoded in a physical state, $\rho(\theta)$, of the system. To achieve this goal, a measurement of elements $\{\Pi_x\}$ is performed on the system.
Let the probability distribution for the measurement outcome $x$ for a given measurement strategy be given by $f(x|\theta)$.
In a single measurement, let us assume that the outcome is $x_1$. Based on this end-result, we have to speculate the value of $\theta$, which is represented by an estimator function given by $g(x_1)$. For a fixed $\theta$, our estimation is correct on an average if we repeat the experiment for large number of times: 
\begin{equation}
    \langle g(x) \rangle_\theta = \int dx f(x|\theta) g(x) = \theta. 
    \label{brisTi}
\end{equation}
Such an estimator $g(x)$, whose average is given by the true value of the estimated parameter is called an unbiased estimator~\cite{Rafao2,review1}. 
Given a \(\theta\), we therefore have a distribution of the estimate \(g(x)\) with 
probability \(f(x|\theta)\). The mean of this distribution is given by the middle term of Eq.~(\ref{brisTi}). We now deal with the variance of the distribution in the case when the estimator is unbiased.

A lower bound on the variance of the estimator function
is given by the Cram\'{e}r-Rao bound~\cite{Maccone,Cramer,Hayashi3}. For an unbiased estimator, the classical Cram\'er-Rao bound  is given by (using the customary notation \(\Delta^2 \theta\) for 
\(\Delta^2 g(x)|_\theta\), the variance of the distribution of the estimate \(g(x)\))
\begin{equation}
\label{ClCramerRao1}
    \Delta^2 \theta \geq \frac{1}{F(\theta)},
\end{equation}
where $F(\theta)$ is the Fisher information (FI)~\cite{review1}, defined as
\begin{eqnarray}
\label{fisher0}
F(\theta) &=& \int dx f(x|\theta) \big[\frac{\partial}{\partial \theta}\log f(x|\theta)\big]^2.
\end{eqnarray}
The base of $\log$ used in the paper is $e$.
The integration here is over the measurement outcomes for a given measurement strategy.
For $\nu$ runs of the measurement \(\{\Pi_x\}\) on the system (with state \(\rho(\theta)\)) with the measurement outcomes $x=\{x_1,x_2,\cdots,x_\nu\}$, 
the Fisher information, $F^{\nu}(\theta)$, takes the following form,
\begin{eqnarray}
&&F^{\nu}(\theta)=
\int dx_1...dx_\nu f(x_1|\theta)...f(x_\nu|\theta)\nonumber\\
&&\phantom{amay dubaili re}\times\big[\frac{\partial}{\partial \theta}\log\big( f(x_1|\theta)...f(x_\nu|\theta)\big)\big]^2.
\end{eqnarray}
It is assumed here that the $\nu$ measurements are independent of each other with the fixed initial conditions.

An important property of the Fisher Information is its additivity. So, if the measurement is repeated $\nu$ times, then
$F^{\nu}(\theta)=\nu F(\theta)$ and inequation~(\ref{ClCramerRao1}) attains the form, 
\begin{equation}
\label{ClCramerRao}
    \Delta^2 \theta \geq \frac{1}{\nu F(\theta)},
\end{equation}
which is the general form of the Cram\'er-Rao bound.

The Cram\'er-Rao bound depends on the measurement strategy that we choose. To make it independent of such a choice, the Fisher information needs to be maximized with respect to all possible measurements in order to get the minimum value for the bound, given in terms of the quantum Fisher information $F_Q(\theta)$. Therefore, we have
\begin{equation}
\label{QCRbound}
\Delta \theta \geq \frac{1}{\sqrt{\nu \big(\max_{\Pi_x} F(\theta)\big)}} = \frac{1}{\sqrt{ \nu F_Q(\theta)}}. 
\end{equation}
This inequality
is usually referred to as
the quantum Cram\'er-Rao bound.
Here $f(x|\theta)=\text{Tr}[\Pi_x \rho(\theta)]$, with $\{ \Pi_x\}$ being a positive operator valued measurement, and $\rho(\theta)$ is a quantum state. $F_Q(\theta)$ is known as the quantum Fisher information which can be expressed as 
\begin{equation}
\label{QFI}
F_Q(\theta)=\text{Tr}\Big[\rho(\theta) {L_s\big[ \rho(\theta) \big]}^2 \Big],
\end{equation}
with $L_s\big[ \rho(\theta) \big]$ being the symmetric logarithmic derivative (SLD)~\cite{Braunstein,Rafao1} and is defined through the expression,
\begin{equation}
    \frac{\partial \rho(\theta)}{\partial \theta}=\frac{1}{2}\Big[L_s \big[ \rho(\theta) \big] \rho(\theta)+\rho(\theta) L_s \big[ \rho(\theta) \big]\Big].
\end{equation}
If the eigenspectrum of $\rho(\theta)=\sum_i \lambda_i(\theta) \ket{e_i(\theta)}\bra{e_i(\theta)}$, then the SLD can be evaluated as
\begin{equation}
\label{SLD}
    L_s \big[ \rho(\theta) \big] = \sum_{i,j}\frac{2 \bra{e_i(\theta)} \frac{\partial \rho(\theta)}{\partial \theta} \ket{e_j(\theta)}}{\lambda_i(\theta)+\lambda_j(\theta)} \ket{e_i(\theta)}\bra{e_j(\theta)},
\end{equation}
where the sum is over pairs \((i,j)\) for which $\lambda_i(\theta)+\lambda_j(\theta)\ne 0$~\cite{Braunstein,Rafao1}. 
The maximum in (\ref{QCRbound}) is attained, for example, for the projective measurement in the eigenbasis of SLD~\cite{Braunstein,Rafao1,Rafao2}. 


In this entire formulation, the initial state, $\rho_0^{\otimes \nu}$, is kept fixed and optimization is made only over the measurement strategy. But one can also consider the best input state. In such a ``two-step optimization" procedure~\cite{Maccone}, the result obtained from (\ref{QCRbound}) is minimized with respect to the input state parameters and we denote 
\begin{equation}
\Delta \theta_{opt}  = \min_{\rho_0^{\otimes\nu}} \frac{1}{\sqrt{ \nu F_Q(\theta)}},
\label{theta_opt}
\end{equation}
and refer to it as the ``optimum \(\Delta \theta\)''.

If the parameter to be estimated is encoded in a pure state \(|\psi(\theta)\rangle\), 
then Eq.~(\ref{QFI})  reduces to
\begin{equation}
    F_Q(\theta)=4 \big[\braket{\dot{\psi}(\theta)|\dot{\psi}(\theta)}-|\braket{\dot{\psi}(\theta)|\psi(\theta)}|^2 \big],
\end{equation}
where the dots represent derivatives with respect to \(\theta\). Suppose now  that there are $\nu$ copies of the input state, where each copy is the $N$-party entangled state, $\ket{\psi^{(N)}_0}$, and this input unitarily evolves through the equation $\ket{\psi^{(N)}_{\theta}}^{\otimes \nu}=U(\theta) \ket{\psi^{(N)}_0}^{\otimes \nu}$, where $U(\theta)$ is a unitary operator given by $U(\theta)=e^{-i\tilde{H}\theta}$, with $\tilde{H}$ being the generator of the unitary having the unit of $\theta^{-1}$.
As $\tilde{H}$ is independent of $\theta$, the quantum Fisher information comes out to be independent of $\theta$ as $F_Q(\theta)=4\Delta^2\tilde{H}$. Hence, the Cram\'er-Rao bound turns out to be
\begin{equation}
    \Delta^2\theta \ge \frac{1}{4\nu\Delta^2\tilde{H}}.
\end{equation}
This inequality tells that $\Delta^2\theta$ can be minimized by maximizing $\Delta^2\tilde{H}$.
It can be shown that the maximum of $\Delta^2\tilde{H}$ can be obtained by choosing the initial state as 
\begin{equation}
\label{best_input}
   \ket{\psi_{0}^{(N)}}=\big( \ket{\lambda_{\max}^{(N)}} + \ket{\lambda_{\min}^{(N)}} \big)/\sqrt{2},
\end{equation}
where $\ket{\lambda_{\max}^{(N)}}$ and $\ket{\lambda_{\min}^{(N)}}$ are the respective eigenvectors associated with the highest and lowest eigenvalues of the generator $\tilde{H}$. Now, let us consider the unitary to be of the form $U(\theta,\phi)=e^{-i(\tilde{H}_1\theta+\tilde{H}_2\phi)}$ and $[\tilde{H}_1,\tilde{H}_2]=0$. For this situation, while estimating $\theta$ and $\phi$ separately, the Cram\'er-Rao bounds for the two parameters are $\Delta^2\theta \ge \frac{1}{4\nu\Delta^2\tilde{H}_1}$ and $\Delta^2\phi \ge \frac{1}{4\nu\Delta^2\tilde{H}_2}$ respectively. Here again, the choice of the best input states are same as in Eq.~(\ref{best_input}), with $\ket{\lambda_{\max}^{(N)}}$ and $\ket{\lambda_{\min}^{(N)}}$ being the eigenvectors respectively associated with the highest and lowest eigenvalues of the generator $\tilde{H}_1$ for estimating $\theta$, and the same of $\tilde{H}_2$ for estimating $\phi$. 
The situation is more involved 
when $\tilde{H}_1$ and $\tilde{H}_2$ do not commute. Then, one cannot write the Cram\'er-Rao bound in terms of the variance of the generator of the unitary and
in such a situation, the best choice of input states cannot be obtained using the  prescription in~(\ref{best_input}).

The phase or the frequency estimation protocol using quantum metrology has been studied in~\cite{Maccone,Cirac&Plenio}. They considered a situation where the objective is to measure an unknown frequency $\omega$. It comes out as a relative phase $\phi(=\omega t)$ taken up by two orthogonal states $|0\rangle$ and $|1\rangle$ when their linear superposition is acted upon by the Hamiltonian,
\begin{equation}
\label{H_0}
H_0=-\hbar\omega \ket{1}\bra{1}.
\end{equation}
In the noiseless scenario, 
a single particle with the initial state $\frac{1}{\sqrt{2}}(|0\rangle+|1\rangle)$ undergoes a time evolution governed by the Hamiltonian $H_0$, and finally the probability $p$ of obtaining the initial state in the final state, after the time evolution, is measured. This probability $p$, for a single particle measurement is given by
\begin{equation}
    p=(1+\cos(\omega t))/2.
\end{equation}
For $n$ uncorrelated qubits, the system evolves by an $n$-qubit Hamiltonian of the form,
\begin{eqnarray}
    \mathcal{H}_n&=&H_0\otimes \mathcal{I}_2\otimes\mathcal{I}_3\cdots \otimes \mathcal{I}_n+\mathcal{I}_1\otimes H_0\otimes\mathcal{I}_3\cdots \otimes \mathcal{I}_n\nonumber\\
    &&\phantom{aji dhaner}+\cdots +\mathcal{I}_1\otimes\mathcal{I}_2\cdots \otimes\mathcal{I}_{n-1} \otimes H_0,
\end{eqnarray}
where each of the qubits is evolving by $H_0$ and $\mathcal{I}_k$ is the identity operator on the k$^{th}$ qubit space. After the evolution, the same  measurement as in the single-qubit case is made on each of the qubits independently, after tracing out the others. 
If the total time of the experiment is $T$, and if every step consists of an evolution for time $t$ and an instantaneous measurement, the total number of measurements will be $\nu=n\frac{T}{t}$.
So, for $n$ copies of uncorrelated single qubits, we can obtain the deviation in the estimation of the frequency $\omega$ as
\begin{equation}
\label{nl_prod}
    \Delta \omega_p = \frac{1}{\sqrt{ntT}},
\end{equation}
when $\nu$ is large. This is often referred to as the shot-noise limit in the literature~\cite{Cirac&Plenio}.
In the maximally entangled input case, where among $n$ qubits, clusters of $N$ qubits are bunched into maximally entangled groups, having the initial state $(|0\rangle ^{\otimes N}+|1\rangle ^{\otimes N})/\sqrt{2}$, the number of measurements for $n$ qubits is $\nu=\frac{n}{N}\frac{T}{t}$. 
Note that we are calling the Greenberger–Horne–Zeilinger state~\cite{Greenberger,Mermin} as the maximally entangled state.
The uncertainty in measuring $\omega$ in this case for large $\nu$ is found to be
\begin{equation}
\label{nl_ent}
    \Delta \omega_e = \frac{1}{\sqrt{ntTN}}.
\end{equation}
Now, comparing Eqs.~(\ref{nl_prod}) and~(\ref{nl_ent}) we can see that a quantum advantage of $\frac{1}{\sqrt{N}}$ is attained for the estimation of $\omega$ in the maximally entangled case over the classical one. 

Next, the same situation is considered but in presence of noise~\cite{Cirac&Plenio,Maccone}, which is inevitably present in any realistic scenario. In presence of dephasing noise, 
for a single-qubit density matrix $\rho$ the dynamical equation can be described by the  Gorini–Kossakowski–Sudarshan–Lindblad master equation given by
\begin{equation}
\label{Lindblad}
    \Dot{\rho} = -\frac{i}{\hbar}\big[H_0,\rho\big] +\frac{\gamma}{2}(\sigma_z \rho \sigma_z - \rho), 
\end{equation}
where $\gamma$ is the decay constant of the dephasing channel having the unit of $\frac{1}{t}$. 
Solving Eq.~(\ref{Lindblad}) for the single-qubit case, and generalizing for $n$ qubits, the expression for minimum $\Delta \omega$ can be obtained ~\cite{Cirac&Plenio}. For the noisy case, it is observed that, for large $\nu$, the minimum values of $\Delta \omega$, viz. $\Delta \omega_{p_{Noise}}$ and $\Delta \omega_{e_{Noise}}$, obtained in the product and maximally entangled scenarios respectively, are the same and hence, both of these settings obtain the same precision in measuring the frequency $\omega$ in presence of decoherence. Thus, while there is an advantage in the maximally entangled initial state in absence of noise, it disappears when a dephasing channel is applied. An important point to be noted here is that while extending from a single-qubit scenario to an $n$-qubit configuration, an underlying assumption is made regarding the qubits' evolution. They evolve through distinct and uncorrelated local dephasing channels, which remain unchanged even in the presence of interactions between the system particles. This assumption underpins the presentation of the total dissipative term within the Lindblad master equation as a summation of independent dissipative terms, originating from the individual qubits experiencing local dephasing channels. This particular circumstance also finds relevance in the context of some setups of thermal devices, where qubits locally interact with thermal reservoirs~\cite{Popescu,transistor}. In such cases, the Lindblad master equation governing the dynamics of the reduced system encompasses a dissipative term that is, essentially, the sum of dissipative terms corresponding to each individual qubit. All of these considerations remain consistent within the framework of the Born-Markov assumptions~\cite{Petruccione, Alicki,Rivas,Lidar}. Within the context of this paper, we rigorously investigate scenarios that adhere to these principles. 
here exist alternate ways of generalizing to multiple qubits~\cite{spinstar,nonmark,nonmark2,spinstar2,heatcurrent,nmfridge}, but we do not discuss those scenarios. 
We adhere to two types of measurement schemes in this paper. These are as follows.
\\

\noindent
\textbf{Scheme I:} The first measurement scheme is a  product measurement having four outcomes. The measurement operators of this four-outcome measurement belong to a set of biorthogonal states, viz. $\{\ket{\Psi_1}\otimes\ket{\Psi_2}, \ket{\Psi_1}\otimes\ket{\Psi_2}^{\perp},\ket{\Psi_1}^{\perp}\otimes\ket{\Psi_2}, \ket{\Psi_1}^{\perp}\otimes\ket{\Psi_2}^{\perp}  \}$. We perform a rank-one projective measurement on the encoded state with the four outcomes corresponding to the projectors onto the biorthogonal vectors.
Let the probability of each of the four measurement outcomes after the measurement be denoted by $p_i$, where $i=1$ to $4$ and $\sum_{i=1}^4 p_i=1$. Once we get the probabilities, we calculate the Fisher information corresponding to the encoded state, using Eq.~\eqref{fisher0}. Finally we optimize over the measurement operators to obtain the best possible metrological precision under this measurement scheme, for the given encoded state. The significance of this type of measurement is that it is locally implementable, and is therefore easier to perform in a realistic situation. We employ the same measurement strategy for both correlated and uncorrelated input probes, in noiseless as well as noisy scenarios, \emph{separately} for non-interacting and interacting encoding Hamiltonians. The results corresponding to this measurement scheme are presented in Sec.~\ref{LO}.\\

\noindent
\textbf{Scheme II:} This measurement scheme is the optimal measurement scheme for a specified encoded state, optimized over arbitrary measurement schemes. This involves a four-outcome projective measurement onto the eigenbasis of SLD of the relevant encoded state.  We evaluate the quantum Fisher information in this case using Eq.~\eqref{QFI}. This measurement strategy is utilized both for uncorrelated and correlated inputs, in noiseless as well as noisy situations, and \emph{separately} for non-interacting and interacting encoding Hamiltonians. We discuss the results with this  measurement in Sec.~\ref{sld0}.\\

An important consideration of our work is that in noisy scenarios, we assume an exact knowledge  of the noise and have the freedom to choose the optimal input state accordingly. Hence in the noisy and noiseless scenarios,  we execute the measurements on different input states, but both of them are optimal in the relevant situations. 
\\

\begin{figure}
\centering
\includegraphics[width=8cm]{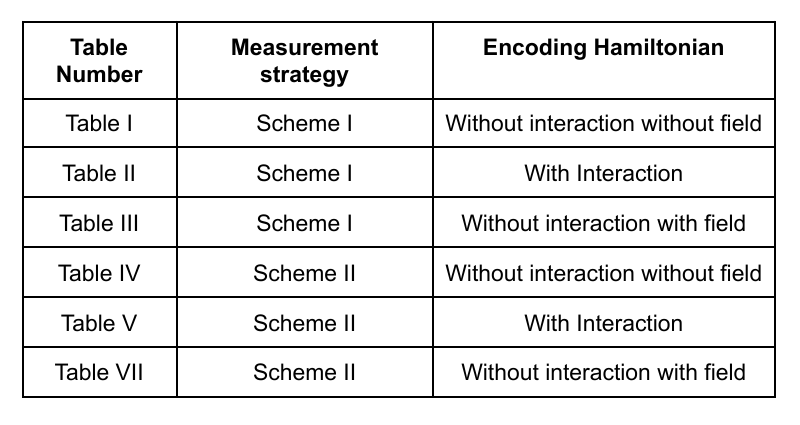}%
\caption{We present here a summary of the cases considered in the different tables in this paper. Table~VI is not included in the above summary, and involves a scaling analysis for a minimum uncertainty for frequency estimation within Scheme II.}
\label{natun}
\end{figure}

The results are presented in several tables in the  two succeeding sections. Fig.~\ref{natun} presents a gist of the cases considered therein.

\section{Parameter estimation using LO measurements (Scheme I)}
\label{LO}
In this section, we consider the Hamiltonian $\mathcal{H}_2$, along with some  modifications in it, and minimize the error in the estimation of a certain parameter of the relevant Hamiltonian.
We restrict to two types of two-qubit pure input probes: arbitrary product and arbitrary maximally entangled ones. The parameter that we want to estimate is encoded onto the input probe via the Hamiltonian. The evolution of the probes is governed by the same Hamiltonian with or without a noisy background.
The noise is modeled by a dephasing channel, and 
is uncorrelated, so that it acts locally on the two subsystems. 
Throughout the paper, we refer to the unitary case  as the noiseless scenario, and the case when dephasing is active  as the noisy one. After the encoding, a suitable measurement is performed on the encoded state. Then we calculate the Fisher information from the measurement outcomes, using Eq.~\eqref{fisher0}. Finally to achieve the best possible accuracy in the relevant setting, we maximize the Fisher information over the choice of input states and measurements. 
 In the rest of the paper, we will call the Fisher information-based lower bound of the minimum deviations of any estimated parameter $\epsilon$ as $\Delta \tilde{\epsilon}$. The evaluation of $\Delta \tilde{\epsilon}$ follows exactly the same method as for obtaining $\Delta\theta_{opt}$, but for finite $\nu$. The minimum of $\Delta\tilde{\epsilon}$ over the evolution time is presented as $\Delta\tilde{\epsilon}_{min}$.
In the optimization processes, the arbitrary initial states chosen for product and maximally entangled cases, as denoted by the subscripts $p$ and $e$ respectively, are given by
\begin{eqnarray}
\label{input}
|\psi_p\rangle &=& \ket{\phi_1} \otimes \ket{\phi_2} \quad \text{and} \nonumber\\
|\psi_e\rangle &=& U_A \otimes U_B  \frac{1}{\sqrt{2}}\big(\ket{00}+\ket{11} \big),
\label{initial}
\end{eqnarray}
where $|\phi_j\rangle=\cos\frac{\theta_j}{2}|0\rangle+e^{i\delta_j} \sin \frac{\theta_j}{2}|1\rangle$ for $j=1,2$ and $U_A,U_B \in SU(2)$ are unitaries acting on the local subsystems. The general form of such a single-qubit unitary is given by
\begin{equation}
    U_k=\left( {\begin{array}{cc}
        e^{i (\alpha_k+\beta_k)/2}\cos{\frac{\gamma_k}{2}} & e^{-i(\alpha_k-\beta_k)/2}\sin{\frac{\gamma_k}{2}} \\
        -e^{i (\alpha_k-\beta_k)/2}\sin{\frac{\gamma_k}{2}} & e^{-i(\alpha_k+\beta_k)/2}\cos{\frac{\gamma_k}{2}} 
        \end{array} } \right),
        \label{unitary}
\end{equation}
for $k=A$ and $B$. Since  $U_A$ and $U_B$ act locally, they can not change the entanglement content of the maximally entangled state chosen. They can at best transform one maximally entangled state into another and one product state into another. Hence, we can optimize over all possible product and, separately, overall maximally entangled states by optimizing over $\theta_j$ and $\delta_j$ in the product input cases for $j=1,2$, and the parameters of the unitaries $U_A$ and $U_B$ for the maximally entangled input cases, to obtain the best choice of initial states for achieving minimum uncertainty of the  parameter to be estimated.

An optimization over the relevant measurements is also required. As elaborated earlier, this paper encompasses two distinct measurement settings. In this section, we deal with the LO measurement performed on two parties.
The measurement operators, as discussed previously are described by a set of biorthogonal states, $\{\ket{\Psi_1}\otimes\ket{\Psi_2}, \ket{\Psi_1}\otimes\ket{\Psi_2}^{\perp},\ket{\Psi_1}^{\perp}\otimes\ket{\Psi_2}, \ket{\Psi_1}^{\perp}\otimes\ket{\Psi_2}^{\perp}  \}$.  The explicit forms of $\{\ket{\Psi_i}$ and $\{\ket{\Psi_i}^{\perp}$ are given by
\begin{eqnarray}
    &&\ket{\Psi_i} = \ket{\phi_{k}} \quad \text{and} \nonumber \\
    &&\ket{\Psi_i}^{\perp} = \sin\frac{\theta_{k}}{2}|0\rangle-e^{i\delta_{k}} \cos \frac{\theta_k}{2}|1\rangle,
\end{eqnarray}
where $k=i+2$ and $i=1,2$.
So we optimize over the parameters $\theta_k$ and $\delta_k$ to obtain the optimal measurement under this type of scenario.  

We do the two-step optimizations, viz., of probes and measurement operators, simultaneously. Whenever numerical optimization is required, we do so using the algorithms of NLOPT~\cite{NLOPT}. In all further discussions regarding precision measurements of parameters, these two optimizations, one over the initial states and the other over the measurements, are taken into account. 

\subsection{Frequency estimation in absence of field and interactions}
\begin{table*}[!htb]
\centering
\begin{tabular}{|c|c|c|c|c|c|}
\hline
   Hamiltonian & \phantom{ami bari}Product\phantom{ami bari} & Maximally entangled & \phantom{ami bari}Product\phantom{ami bari} & Maximally entangled\\ (Estimated parameters) &  (noiseless) &  (noiseless) &  (noisy) &  (noisy) \\ 
  \hline
  \hline
  $\mathcal{H}_2$ ($\omega$) & 0.35 & 0.25 & 0.82 & 0.82 \\ 
   \hline
\end{tabular}
\caption{Minimum uncertainty obtained from Fisher information-based lower bound $\tilde{T}\Delta \tilde{\omega}_{min}$ for the estimated parameter $\omega$ for the ``ideal" Hamiltonian $\mathcal{H}_2$. Here the optimization is over initial states, measurement strategies within Scheme I, and evolution time, for $N=2$, $n=2$, $T/\tilde{T}=2.0$, $\tilde{T}\omega=0.5 \times \pi \times 10$. The numerical values are correct up to the second decimal place.
All $\tilde{T}\Delta\tilde{\omega}_{min}$ are obtained for the time interval $0$ to $2\tilde{T}$. All numbers in the table are for dimensionless quantities.}
\label{lo_w}
\end{table*}

We begin by considering the encoding Hamiltonian given in Eq.~\eqref{H_0}, where the parameter, $\omega$, is to be estimated. The Fisher information-based lower bound is evaluated in the estimation of $\omega$. In Table~\ref{lo_w}, the dimensionless quantity $\tilde{T}\Delta\tilde{\omega}_{\text{min}}$, is given, which is being measured by optimizing the initial states, measurement and the dimensionless time, $t/\tilde{T}$.  Here $\tilde{T}$ is a constant having the unit of time. The investigations are done for $n=2$.
From Table~\ref{lo_w}, we find that there is an advantage of using an initial maximally entangled state over the initial product state in estimating the frequency in noiseless scenario.
This advantage vanishes in presence of uncorrelated local dephasing noise. These results presented in Table~\ref{lo_w}, both in the noiseless and noisy cases, pertain to the measurement Scheme I as mentioned previously. A similar observation was given in ~\cite{Cirac&Plenio}. However, the measurement strategy employed by them is different from ours.
Let us briefly state their measurement scheme. Suppose that a single copy of the initial probes is $\ket{\psi_0}$. They construct a set of measurement elements containing two rank-one projectors for uncorrelated (qubit) inputs, and a rank-one projector and a rank-$(2^N-1)$ projector for copies of an $N$-qubit GHZ state as inputs,  viz. $\big\{\ket{\psi_0}\bra{\psi_0},\mathcal{I}-\ket{\psi_0}\bra{\psi_0}\big\}$, and perform projective measurements on each of the encoded states using this strategy, where \(\mathcal{I}\) is the identity operator on the Hilbert space of a single copy of the probes. This furnishes the probability of getting - or not - the initial state at the output. 
They have considered this measurement strategy in noiseless scenario with the initial state being either a product state or - separately - a maximally entangled (GHZ) state. In the noisy situation, they repeat the same projective measurement on the same initial probes $\ket{\psi_0}$, as in the noiseless case. This is e.g. the useful choice when the presence of noise is unknown to the observers.

In practical applications, therefore, the prevalence of noise in natural systems limits the quantum advantage of employing a maximally entangled initial state for frequency estimation. Notably, this limitation is particularly pronounced in scenarios involving uncorrelated local dephasing noise, whereas the implications for generic dephasing channels remain subject to further study. We now aim to find whether any quantum advantage can be achieved in estimating some system parameters by altering the system Hamiltonian, possibly by incorporating transverse fields or/and interactions between the particles involved. To fulfill this goal, we consider several single qubit fields and two-qubit interactions between the system particles and evaluate the minimum uncertainty in estimating the frequency and other field and coupling parameters by utilizing optimum initial states and optimal measurement strategies. For the measurement strategy, we, however, pertain to optimum LO measurement given by scheme I.

\subsection{Ising coupling incorporation for restoration of quantum metrological advantage}

\begin{table*}[!htb]
\centering
\begin{tabular}{|c|c|c|c|c|c|}
\hline
   Hamiltonian & \phantom{ami bari}Product\phantom{ami bari} & Maximally entangled & \phantom{ami bari}Product\phantom{ami bari} & Maximally entangled\\ (Estimated parameters) &  (noiseless) &  (noiseless) &  (noisy) &  (noisy) \\ 
  \hline
  \hline  
  $H_1$ ($\omega$) & 0.35 & 0.25 & 0.95 & 0.82 \\
  \hline 
  $H_2$ ($\omega$) & 0.35 & 0.25 & 0.95 & 0.82 \\ 
  \hline
  $H_1$ ($J$) & 0.25 & 0.25 & 0.48 & 0.54 \\
  \hline
  $H_2$ ($J$) & 0.25 & 0.25 & 0.48 & 0.54 \\
  \hline
  $H_3$ ($J$) & 0.25 & 0.25 & 0.48 & 0.54 \\
  \hline
  $H_2$ ($h$) & 0.80 & 0.58 & 0.79 & 0.58 \\
  \hline
\end{tabular}
\caption{Minimum uncertainty obtained from Fisher information-based lower bound $\tilde{T}\Delta \tilde{\epsilon}_{min}$ for an estimated parameter $\epsilon$ for various Hamiltonians observed by incorporating two-body interaction terms in the ideal Hamiltonian $\mathcal{H}_2$.  Here the optimization is over initial states, measurement strategies within Scheme I, and evolution time, for $\tilde{T}J=0.5$, $\tilde{T}h=0.5$ and $\tilde{T}\gamma=0.5$.  The numerical values are correct up to the second decimal place. All numbers in the table are for dimensionless quantities.}
\label{lo_J}
\end{table*}

In this subsection, we investigate whether the advantage of using maximally entangled initial states in the noiseless case remains also in a noisy environment in presence of two-qubit  interactions between the system particles in presence and/or absence of transverse fields. In Table~\ref{lo_J}, we have presented the optimum lower bounds of the deviations of estimated parameters which are being measured by optimizing the initial states, measurement and the evolution time, in the cases where the Hamiltonian possesses such a coupling term.

\subsubsection{Estimation of frequency}

\begin{figure*}
\centering
\includegraphics[width=8cm]{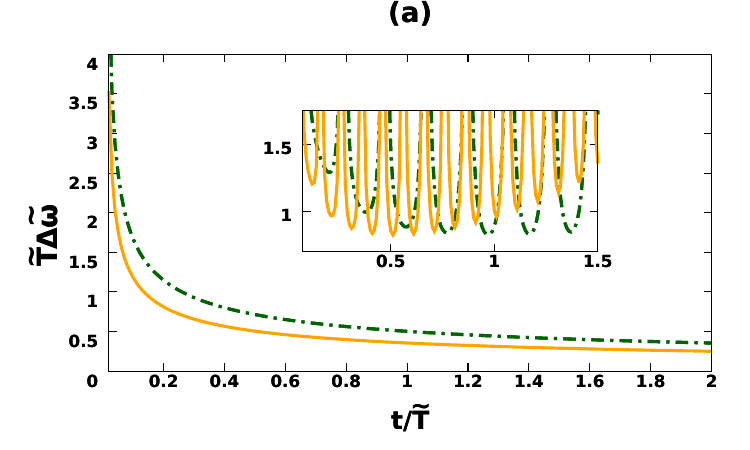}%
\hspace{.25cm}%
\includegraphics[width=8cm]{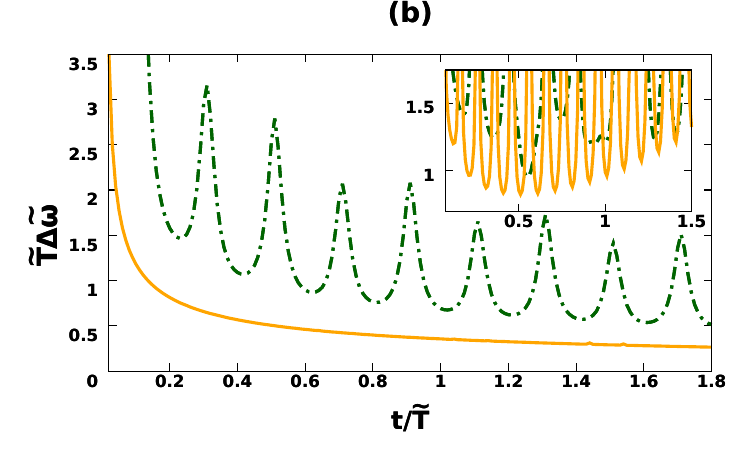}%
\caption{Fisher information-based lower bound of the minimum uncertainty in frequency measurement. Here we have depicted the time dynamics of $\tilde{T}\Delta \tilde{\omega}$ for both product and maximally entangled initial states under the evolution of the Hamiltonian $H_1$ and $H_2$ in the panels (a) and (b) respectively, while considering measurement Scheme I. The figures correspond to noiseless situations and insets depict the noisy scenarios. All other considerations are the same as in Table~\ref{lo_J}. The solid orange curves correspond to the optimal choice of maximally entangled initial state and the dotted-dashed green curves represent the same for the product initial state. All quantities plotted are dimensionless.}
\label{int_noise_w}
\end{figure*}

The minimum error in estimation of the frequency, \(\omega\), present in the encoding interacting Hamiltonians, is given in this subsection. At first, we modify the ideal Hamiltonian $\mathcal{H}_2$ by introducing an Ising interaction between the particles involved. The total Hamiltonian of the system therefore looks like
\begin{equation}
\label{H1}
    H_1=\mathcal{H}_2+\hbar J(\sigma_z^1 \otimes \sigma_z^2),
\end{equation}
with $J$ being a coupling constant having the unit of $\text{time}^{-1}$. 

Let us look into the calculation of the Fisher information-based lower bound from the joint probability distributions of each outcome obtained after measurement using Scheme I.
We consider the situation of initiating with product input probes, and where the evolution is considered in the presence of 
a noisy environment. The initial state, $\ketbra{\psi_p}{\psi_p}$, evolves to some final state, say $\rho_f$, in presence of local dephasing noise.
A four-outcome LO measurement is performed on the evolved state, and the probability of obtaining each outcome after the measurement, denoted by $p_i$ for $i=1$ to $4$, is given by
\begin{eqnarray}
\label{p}
p_1 &=& \bra{\Psi_1 \Psi_2}\rho_f\ket{\Psi_1 \Psi_2} \nonumber \\
p_2 &=& \bra{\Psi_1 \Psi_2^{\perp}}\rho_f\ket{\Psi_1 \Psi_2^{\perp}} \nonumber \\
p_3 &=& \bra{\Psi_1^{\perp} \Psi_2}\rho_f\ket{\Psi_1^{\perp} \Psi_2} \nonumber \\
p_4 &=& \bra{\Psi_1^{\perp} \Psi_2^{\perp}}\rho_f\ket{\Psi_1^{\perp} \Psi_2^{\perp}}.  
\end{eqnarray}
The expressions of $p_i$, for $i=1$ to $4$ are given in Appendix~\ref{ap1}. Using these probabilities, we evaluate the Fisher information according to Eq.~\eqref{fisher0}, to obtain the deviations in the estimations of frequency. The deviations, thus obtained, are optimized over the choice of input state, measurement strategy under Scheme I and the evolution time, whereby we reach the quantity, $\Delta\tilde{\omega}_{min}$.

A similar analysis is carried out for maximally entangled initial probes. The expressions of the corresponding probabilities in this case are not explicitly presented in the paper, as they are long, and as they can be calculated in the same manner as for the product input case.
For all the other scenarios considered in the paper, the minimum deviations in the estimations of the relevant parameters have been computed analogously, for product and maximally entangled input probes.

The minimum error in the estimation of $\omega$, given by $\tilde{T}\Delta \tilde{\omega}_{min}$, is depicted in Table~\ref{lo_J}. We observe that in the noiseless scenario, the 
situation is similar to the case without interactions, i.e., 
optimum maximally entangled probes are advantageous over optimum product ones. However, in the noisy scenario, there is a restoration of metrological advantage by using optimal maximally entangled input over optimal product ones, in the estimation of the minimum error in $\omega$. 
If we look at the time dynamics of $\tilde{T}\Delta \tilde{\omega}$  depicted in Fig.~\ref{int_noise_w}-(a) for $H_1$, we observe that, in noiseless scenarios, the quantity, $\tilde{T}\Delta \tilde{\omega}$, is hyperbolic in nature for both product and maximally entangled inputs in the noiseless case. In contrast, under noisy scenarios, $\tilde{T}\Delta \tilde{\omega}$ is oscillatory for both product and maximally entangled inputs.

\begin{figure}
\centering
\includegraphics[width=8cm]{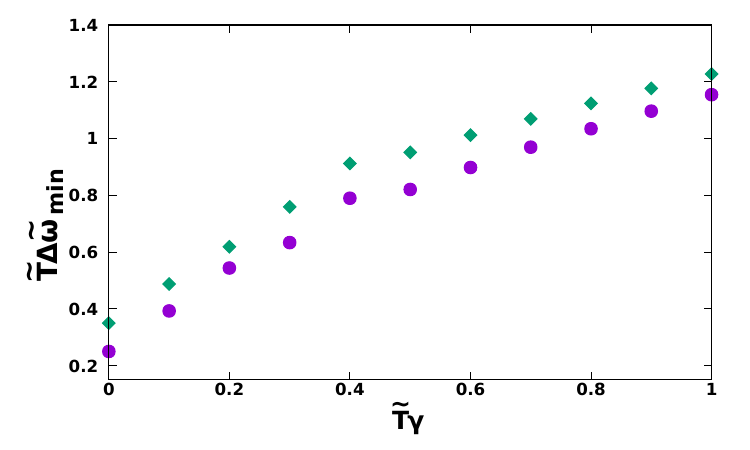}%
\caption{Variation of the Fisher information-based lower bound of the minimum uncertainty in the estimation of frequency with the noise strength $\gamma$. Here we depict the quantity $\tilde{T}\Delta \Tilde{\omega}_{min}$ with the increase of the noise strength $\gamma$ for the Hamiltonian $H_1$. The optimization is over initial states, measurement strategy using Scheme I and evolution time. The purple circles depict the case of optimum maximally entangled inputs and the  green triangles depict that of the optimum product inputs. All the other parameters are chosen same as in Fig.~\ref{int_noise_w}. The quantities plotted along both the axes are dimensionless.
}
\label{gamma}
\end{figure}

The noise strength of the dephasing channels, denoted by $\gamma$ (see Eq.~(\ref{Lindblad})), can have significant impacts on the measurement precision of the parameters. To investigate the effects of the noise strength on the minimum error of the estimated parameters, we take an instance of measuring frequency using Scheme I for the Hamiltonian $H_1$ (see Eq.~\eqref{H1}). In Fig.~\ref{gamma}, the behavior of the quantity $\tilde{T}\Delta\tilde{\omega}_{min}$ is depicted as the noise strength $\gamma$ increases from $0$ to $1$. It is evident from the figure that the minimum error of the estimation of frequency increases monotonically with the increase of  noise strength for both product and maximally entangled optimum initial states. This trend indicates a degradation in the precision of estimated frequency as noise strength increases. Additionally, the figure illustrates that across the range $0\le \tilde{T}\gamma \le 1$, the measurement precision associated with maximally entangled inputs consistently outperforms that of product inputs, highlighting the advantage of employing maximally entangled initial states.

We now introduce a field term, applied in the direction perpendicular to that in the ideal Hamiltonian $\mathcal{H}_2$. Here we consider both the transverse field and the Ising interaction terms together, while still evaluating the minimum deviation in the estimation of $\omega$. Therefore, the Hamiltonian of the system takes the form
\begin{equation}
    H_2=\mathcal{H}_2+\hbar J\big(\sigma_z^1\otimes \sigma_z^2\big)+\hbar h(\sigma_x^1 + \sigma_x^2).
\end{equation}
The combined effect of the Ising interaction and transverse field is qualitatively the same as that of the Ising interaction
alone, i.e., optimal maximally entangled probe proves to be beneficial over optimal product ones, both in absence and presence of noise. This result is depicted in the second row of Table~\ref{lo_J}.
If we see the time dynamics of $\tilde{T}\Delta \tilde{\omega}$ given in Fig.~\ref{int_noise_w}-(b) for $H_2$, we find that, in noiseless scenarios, the quantity, $\tilde{T}\Delta \tilde{\omega}$, shows a hyperbolic behavior for maximally entangled inputs, while it exhibits oscillatory behavior for product inputs. In contrast, under noisy conditions, the behavior of  $\tilde{T}\Delta \tilde{\omega}$ is oscillatory for both product and maximally entangled inputs. The advantage of using optimal maximally entangled input over the product input in both noiseless and noisy scenarios is also evident from the plot. 

The reason behind the restoration of quantum advantages by introducing an interaction in the Hamiltonian of the system can be comprehended as follows.  In quantum metrology, the initial entanglement of the probes plays a crucial role, giving quantum systems an advantage over classical ones, especially in noiseless situations~\cite{Cirac&Plenio}. Hence, generating entanglement between the initial probes, may be helpful to achieve better precision in estimating system parameters. Also, non-interacting entangled inputs experience entanglement decay over time, and non-interacting initial product inputs remain unentangled throughout. We introduce interactions between system particles in the system's Hamiltonian, as it can lead to entanglement growth over time in certain cases, which in turn can potentially enhance the measurement precision of system parameters.

\subsubsection{Estimation of coupling constant}

Now we consider the estimation of the coupling constant. 
In contrast to the previous subsections, estimating the coupling constant $J$ reveals some significant differences in the Fisher information-based lower bound of measurement precision. We analyze the quantity, $\tilde{T}\Delta \tilde{J}_{min}$, to estimate the precision in measuring the coupling constant, $J$. From the third row of Table~\ref{lo_J}, we observe a stark contrast in the results, compared to the preceding subsections, for the Hamiltonians $H_1$ and $H_2$. In the noiseless scenario, the minimum error in estimating $J$ remains the same for both product and maximally entangled inputs. However, there is an advantage observed for optimal product inputs over optimal maximally entangled inputs over in the noisy scenario.

\begin{figure*}
\centering
\includegraphics[width=8cm]{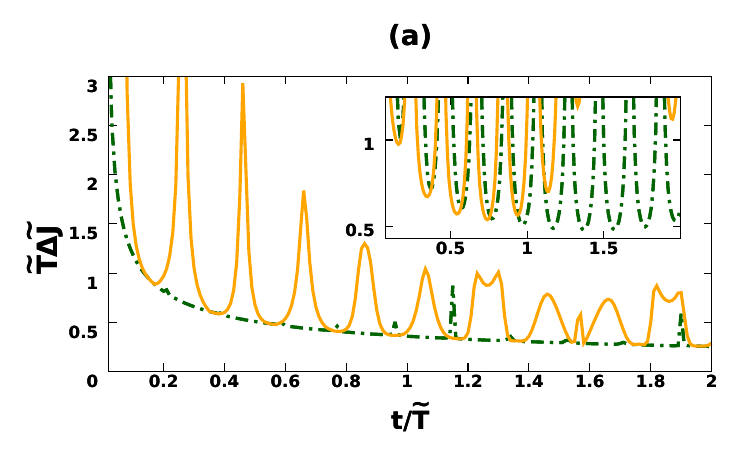}%
\hspace{.25cm}%
\includegraphics[width=8cm]{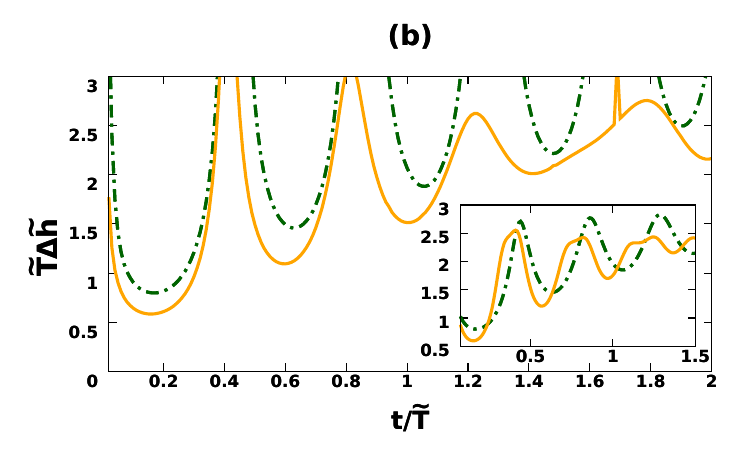}%
\caption{Fisher information-based lower bound of the minimum uncertainty in frequency measurement. Here we have depicted the time dynamics of $\tilde{T}\Delta \tilde{J}$ and $\tilde{T}\Delta \tilde{h}$ for both product and maximally entangled initial states under the evolution of the Hamiltonian $H_1$ and $H_2$ respectively in panels  (a)  and (b). The measurement strategies pertain to scheme I. The figures represent the noiseless cases and insets depict the respective noisy scenarios. All other considerations are the same as in Table~\ref{lo_J}. The solid orange curves correspond to the optimal choice of maximally entangled initial state and the dotted-dashed green curves represent the same for the product initial state. All quantities plotted are dimensionless.}
\label{J_h}
\end{figure*}

For the Hamiltonian $H_1$, the nature of $\tilde{T}\Delta \tilde{J}$ vs $t/\tilde{T}$ in the noiseless scenario looks completely different from the corresponding curves for the case of frequency estimation discussed above. In the time dynamics of $\tilde{T}\Delta \tilde{J}$, there occurs oscillations for maximally entangled inputs whose amplitude decreases with time and for product inputs, the same quantity depicts a hyperbolic behavior. The profile of $\tilde{T}\Delta \tilde{J}$ in the noisy situation exhibits oscillations for both product and maximally entangled initial probes. The product initial state achieves exactly the same precision as that of the maximally entangled one in noiseless scenario. In contrast, in the noisy case, the optimal product probe provides better precision. Refer to Fig.~\ref{J_h}-(a).

We now investigate another Hamiltonian $H_3$ to see whether similar features in the estimation of $J$, which appears in case of $H_1$ and $H_2$,  are also prevalent if we consider $H_3$.
The Hamiltonian $H_3$ is given by
\begin{equation}
    H_3=\hbar J(\sigma_z^1 \otimes \sigma_z^2).
\end{equation}
For this Hamiltonian also, as we can observe from the fifth row of Table~\ref{lo_J} that the optimal product input probes outperform the precision obtainable by optimal maximally entangled initial probes.
So there exists a general trend that product initial states give equal or better precision as that of the maximally entangled inputs in the case of estimation of $J$, in the noiseless and noisy settings respectively, unlike the case of estimation $\omega$. For frequency estimation, on the other hand, there is always a quantum advantage, both in the noiseless and noisy scenarios.

\subsubsection{Estimation of transverse field strength}

We now estimate and analyze the Fisher information-based lower bound of the uncertainty in measuring the field strength $h$ ($\Delta \tilde{h}$) in the same manner as in the previous case of measuring $\omega$ and $J$. For the system Hamiltonians $H_2$, the estimation results, $\tilde{T}\Delta \tilde{h}_{min}$, is given in the sixth row of Table~\ref{lo_J} for Scheme I, for certain instances of the system parameters. 
In both the noiseless and noisy situations, there are quantum advantages if we initiate with the optimal maximally entangled input instead of the optimal product input. 

The behavior of $\tilde{T}\Delta \tilde{h}$ with respect to time $t/\tilde{T}$ for the system Hamiltonian $H_2$, for Scheme I is demonstrated in Fig.~\ref{J_h}-(b). It is evident from the figure that there are quantum advantages both in the noiseless and noisy cases, i.e. there is an enhancement of metrological precision if we initiate with maximally entangled states, whether or not noise is present. In both the noiseless and noisy scenarios, $\tilde{T}\Delta \tilde{h}$ has an oscillatory nature for product and maximally entangled cases, and the amplitude of oscillation decays with the increase of time. 

\subsection{Can quantum metrological advantage be restored in absence of interactions?}

\begin{table*}[!htb]
\centering
\begin{tabular}{|c|c|c|c|c|c|}
\hline
   Hamiltonian & \phantom{ami bari}Product\phantom{ami bari} & Maximally entangled & \phantom{ami bari}Product\phantom{ami bari} & Maximally entangled\\ (Estimated parameters) &  (noiseless) &  (noiseless) &  (noisy) &  (noisy) \\ 
  \hline
  \hline  
  $H_4$ ($\omega$) & 0.35 & 0.25 & 0.82 & 0.82 \\
  \hline 
  $H_4$ ($h$) & 0.82 & 0.58 & 0.82 & 0.58 \\ 
  \hline
\end{tabular}
\caption{Minimum uncertainty obtained from Fisher information-based lower bound $\tilde{T}\Delta \tilde{\epsilon}_{min}$ for an estimated parameter $\epsilon$ for various Hamiltonians observed in absence of interaction  terms in the ideal Hamiltonian $\mathcal{H}_2$. Measurement scheme I is considered here. The numerical values are correct up to the second decimal place. All numbers in the table are for dimensionless quantities.}
\label{lo_h}
\end{table*}

Here we alter the ideal Hamiltonian by adding a transverse field of field strength, $h$.
The total Hamiltonian of the system is therefore
\begin{equation}
    H_4=\mathcal{H}_2+\hbar h(\sigma_x^{1} + \sigma_x^2),
\end{equation}
with $h$ being in unit of time$^{-1}$. In this subsection, we consider the minimum deviations in the estimations of the parameters, $\omega$ or $h$, present in the encoding Hamiltonian, $H_4$, in absence of any two-qubit interactions, while still adhering to measurement Scheme I. The results obtained in this case are depicted in Table~\ref{lo_h}. The quantity, $\tilde{T}\Delta \tilde{\omega}_{min}$, is presented in the first row of the  table.  Referring to this, we find that it is beneficial to consider optimal maximally entangled initial probes than optimal product initial probes in the noiseless scenario. However, in the noisy scenario, both optimal maximally entangled input and optimal product input provides exactly the same precision in the estimation of $\omega$. Therefore, the quantum metrological advantage cannot be restored in absence of interaction terms in the encoding Hamiltonian with optimal maximally entangled inputs in the noisy situation, while estimating the frequency.

Next we consider the estimation of the field strength, $h$, for the same Hamiltonian, $H_4$. Refer to the second row of Table~\ref{lo_h}, where the quantity, $\tilde{T}\Delta \tilde{h}_{min}$, is presented. In this situation, we find that the optimal maximally entangled probes are advantageous over optimal product input probes, both in the noiseless and noisy settings. An important observation here is that the noise has practically no effect on the evaluation of the minimum deviation in the estimation of the transverse field strength.

\section{Parameter estimation with measurement in the eigenbasis of SLD (Scheme II)} \label{sld0}
Though the LO measurement scheme is easier to realize physically, it is always useful to compare its precision with the optimal measurement scheme, which is a projective measurement in the eigenbasis of SLD.
In this section, therefore, we extend our analysis by comparing the outcomes of the previous strategy (Scheme I) with those obtained using the optimal measurement scheme, i.e., the measurement in the eigenbasis of SLD (Scheme II).  Here also, the initial states for uncorrelated and maximally entangled probes, as presented in Eq.~(\ref{initial}), are employed. 
After the measurement using Scheme II, we evaluate the quantum Cram\'{e}r-Rao bound utilizing the quantum Fisher information provided by Eq.~(\ref{QFI}). In particular, our aim is to find whether any quantum metrological advantage can be restored in the noisy scenario, which vanished in the case of encoding Hamiltonian $\mathcal{H}_2$ while adhering to measurement Scheme I, if we implement the optimal measurement strategy.

\subsection{Frequency estimation in absence of field and interactions}
\begin{table*}[!htb]
\centering
\begin{tabular}{|c|c|c|c|c|c|}
\hline
   Hamiltonian & \phantom{ami bari}Product\phantom{ami bari} & Maximally entangled & \phantom{ami bari}Product\phantom{ami bari} & Maximally entangled\\ (Estimated parameters) &  (noiseless) &  (noiseless) &  (noisy) &  (noisy) \\ 
  \hline
  \hline
  $\mathcal{H}_2$ ($\omega$) & 0.35 & 0.25 & 0.82 & 0.80 \\ 
   \hline
\end{tabular}
\caption{Minimum uncertainty obtained from Fisher information-based lower bound $\tilde{T}\Delta \tilde{\omega}_{min}$ for the estimated parameter $\omega$ for the ideal Hamiltonian $\mathcal{H}_2$. Here the optimization is over initial states, measurement strategies within Scheme II, and evolution time, for $N=2$, $n=2$, $T/\tilde{T}=2.0$, $\tilde{T}\omega=0.5 \times \pi \times 10$. The numerical values are correct up to the second decimal place.
All $\tilde{T}\Delta\tilde{\omega}_{min}$ are obtained for the time interval $0$ to $2\tilde{T}$. All numbers in the table are for dimensionless quantities.}
\label{sld_w}
\end{table*}

Let us begin by considering the ideal Hamiltonian $\mathcal{H}_2$.
The Fisher information-based lower bound is evaluated in the estimation of $\omega$. In Table~\ref{sld_w}, the dimensionless quantity $\tilde{T}\Delta\tilde{\omega}_{\text{min}}$, is presented, which is being calculated by optimizing over the initial states, measurement and the dimensionless time, $t/\tilde{T}$. The investigations are done for $n=2$.
Upon examining Table~\ref{sld_w}, it is evident that a minor relative quantum advantage is obtained in noisy scenarios when compared to the ideal cases of the frequency estimation protocol detailed in the  Table~\ref{lo_w}. So, it is intriguing to find that the optimal measurement scheme itself, in absence of any field or interaction terms in the encoding Hamiltonian, favors the revival of quantum advantage which disappeared for measurement scheme I. As previously mentioned, the SLD measurement scheme represents the optimal approach, and we can see that, its application results in a notable reduction in the minimum error for frequency estimation, particularly for optimal maximally entangled inputs as opposed to optimal product inputs in the noisy scenario. Hence, there is restoration of quantum advantage in frequency estimation for the ideal Hamiltonian $\mathcal{H}_2$ in the noisy setting.   
Interestingly, in the ideal scenarios with the same initial probes, both the  measurement scheme I, as well as the SLD measurement scheme, yield identical minimum frequency estimation errors in noiseless situations.

\subsection{Incorporation of Ising coupling to achieve metrological precision in parameter estimation}

\begin{table*}[!htb]
\centering
\begin{tabular}{|c|c|c|c|c|c|}
\hline
   Hamiltonian & \phantom{ami bari}Product\phantom{ami bari} & Maximally entangled & \phantom{ami bari}Product\phantom{ami bari} & Maximally entangled\\ (Estimated parameters) &  (noiseless) &  (noiseless) &  (noisy) &  (noisy) \\ 
  \hline
  \hline  
  $H_1$ ($\omega$) & 0.35 & 0.25 & 0.82 & 0.80 \\
  \hline 
  $H_2$ ($\omega$) & 0.35 & 0.25 & 0.82 & 0.80 \\ 
  \hline
  $H_1$ ($J$) & 0.25 & 0.25 & 0.47 & 0.47 \\
  \hline
  $H_2$ ($J$) & 0.25 & 0.25 & 0.47 & 0.47 \\
  \hline
  $H_3$ ($J$) & 0.25 & 0.25 & 0.47 & 0.47 \\
  \hline
  $H_2$ ($h$) & 0.80 & 0.58 & 0.79 & 0.58 \\
  \hline
\end{tabular}
\caption{Minimum uncertainty obtained from Fisher information-based lower bound $\tilde{T}\Delta \tilde{\epsilon}_{min}$ for an estimated parameter $\epsilon$ for various Hamiltonians observed by incorporating two-body interaction terms in the ideal Hamiltonian $\mathcal{H}_2$. The measurement strategy is according to scheme II. The numerical values are correct up to the second decimal place. All numbers in the table are for dimensionless quantities.}
\label{sld_J}
\end{table*}

In this subsection, we investigate whether the advantage of using maximally entangled initial states in both the noiseless and noisy cases  also persist in presence of two-qubit  interactions and/or transverse fields between the system particles.

\subsubsection{Estimation of frequency}

We here shift our focus to the modified Hamiltonians  $H_1$ and $H_2$, where we assess the minimal error in frequency estimation. The outcomes of this analysis are presented respectively in the first and second rows of Table~\ref{sld_J}. Notably, the obtained values are quantitatively indistinguishable up to the second decimal places when compared with the ideal SLD measurement scenario detailed in Table~\ref{sld_w}. So, in this scenario as well, the quantum advantages are visible both in the noiseless and noisy scenarios.  
Thus, with both the ideal Hamiltonian $\mathcal{H}_2$ and the modified Hamiltonian $H_1$, it is plausible to restore the quantum advantage in noisy scenarios by using the SLD measurement scheme when compared to the ideal case evaluated by using Scheme I, but the degree of restoration is relatively small.

We will now explore scenarios in which we consider a single copy of initial probes, each represented as an $N$-qubit state with $N$ being an arbitrary integer. In these cases also, we specifically focus on two-body interactions between the qubits. Therefore, the Hamiltonian for a single $N$-qubit probe can be expressed as
\begin{equation}
\label{ham_N}
    H_1^{N}=\mathcal{H}_N+\sum_{k=1}^N \hbar J(\sigma_z^{k} \otimes \sigma_z^{k+1}),
\end{equation}
with periodic boundary conditions imposed.
We will now perform the same task as previously undertaken for two-qubit initial probes, as discussed earlier, but this time with $N$-qubit product and entangled inputs, and encoding by a two-body interacting encoding Hamiltonian given in Eq.\eqref{ham_N}, followed by the optimum (SLD) measurement on the encoded probe. An arbitrary $N$-qubit product input state is given by 
\begin{equation}
    \ket{\chi_p^N}= \ket{\phi_1} \otimes \ket{\phi_2} ... \otimes \ket{\phi_N}.
\end{equation}
Here, to obtain the optimal input state, the optimization is performed over all the parameters of $\ket{\phi_1}$, $\ket{\phi_2}$, $\cdots$, $\phi_N$.
For $N$-qubit entangled inputs, we consider the $N$-qubit Greenberger-Horne-Zeilinger (GHZ) states, which are represented as $\ket{\text{GHZ}}=\frac{1}{\sqrt{2}}\big(\ket{0}^{\otimes N}+\ket{1}^{\otimes N}\big)$. Subsequently, we apply local unitary transformations to all the qubits, to cover the entire range of $GHZ$ states expressed in different bases, as
\begin{equation}
    \ket{\chi_e^N}=U_1\otimes U_2 \otimes ... \otimes U_N \frac{1}{\sqrt{2}}\big(\ket{0}^{\otimes N}+\ket{1}^{\otimes N}\big).
\end{equation}
To attain the optimal input state in this scenario, the optimization should encompass all the parameters of $U_1$, $U_2$, $\cdots$, $U_N$.

\begin{table}[!htb]
\centering
\begin{tabular}{|c|c|c|}
\hline
   \;\;\; $N$ \;\;\; & \;\;\; $\tilde{T}\Delta\tilde{\omega}_{min}^P$ \;\;\; &  \;\;\;$\tilde{T}\Delta\tilde{\omega}_{min}^E$ \;\;\;\\  
  \hline
  \hline
  2 & 0.907986 & 0.817733\\ 
   \hline
   3 & 0.742497 & 0.649712  \\ 
   \hline
  4 & 0.643622 & 0.55072\\ 
   \hline
   5 & 0.575897 & 0.490527 \\ 
   \hline
   \end{tabular}
\caption{Minimum uncertainty obtained from Fisher information-based lower bound, $\tilde{T}\Delta \tilde{\omega}_{min}^P$ and $\tilde{T}\Delta \tilde{\omega}_{min}^E$, for the estimated parameter $\omega$ corresponding to optimal product and maximally entangled inputs respectively. Here the optimization is over initial states, measurement strategies within Scheme II, and time. $N$ denotes the number of subsystems involved. All numbers in the table are for dimensionless quantities.}
\label{N}
\end{table}

We estimate the parameter, $\omega$, and find how the minimum deviations in the estimation of $\omega$ scale with $N$ for each of the optimal product and maximally entangled cases. To do this, let us consider $N=N_k$. The minimum error in the estimation of $\omega$ corresponding to $N_k$ parties is denoted by $\Delta \omega_k$. Therefore, $\Delta \omega_k \propto 1/N{_k}^x$, and the ratio of the two minimum deviations corresponding to $k$ and $k+1$ is given by $\Delta \omega_k/\Delta \omega_{k+1}=\left(N_{k+1}/N_K\right)^x$, where we want to find the value of $x$. The value of $x$ can be obtained from the relation 
\begin{equation}
    x=\frac{\ln\left(\frac{\Delta \omega_k}{\Delta \omega_{k+1}}\right)}{\ln\left(\frac{N_{k+1}}{N_k}\right)},
\end{equation}
where $\ln$ denotes the natural logarithm.
We use this relation to find how the quantity, $\tilde{T}\Delta\tilde{\omega}_{min}$, scales with $N$, for each of product and maximally entangled scenarios.
The minimum standard deviations obtained in the estimations of frequencies, corresponding to optimal product and maximally entangled inputs, denoted respectively by $\tilde{T}\Delta \tilde{\omega}_{min}^P$ and $\tilde{T}\Delta \tilde{\omega}_{min}^E$, for different values of $N$, are provided in Table~\ref{N}.
Our findings reveal that the minimum precision in the estimation of $\omega$ scales as $N^{-x}$ with \(x=0.5\) when employing optimal product inputs, and as $1/N^{-x}$ with \(x=0.57\) when using optimal GHZ inputs. Hence, the use of maximally entangled initial probes proves to be more advantageous than the utilization of product inputs, even in the context of $N$-qubit probes, for enhancing measurement precision in frequency estimation.
Here, the multiqubit state is considered as having maximal entanglement in the sense of possessing maximal generalized geometric measure~\cite{ggmshimony,ggmlinden,ggmgoldbart,ggmblasone,ggm0,ggm}.

\subsubsection{Estimation of coupling constant}
Let us now turn our attention to the inclusion of interaction terms in the encoding Hamiltonian, with or without transverse fields,  where we aim to estimate the strength of the coupling constant denoted by $J$. The result is intriguing and distinct from the others.  We observe from the third, fourth and fifth rows of Table~\ref{sld_J} that the optimal entangled input always gives the same precision as the optimal product input, both in the noiseless and noisy scenarios, hence offering no quantum advantages. So, the presence of entanglement in the input probe is not necessary to attain ultimate metrological precision in the estimation of the coupling constant, $J$, for encoding Hamiltonians, $H_1$, $H_2$ and $H_3$.
This trend is different from the relevant situations where we considered measurement scheme I.

\subsubsection{Estimation of transverse field strength}
In the last row of Table~\ref{sld_J}, we consider the Hamiltonain $H_2$, and estimate the field strength $h$. Similar to the patterns observed in the preceding section,  the optimal maximally entangled inputs yield better  precision levels than the optimal product inputs in both the noiseless and noisy situations, as depicted in the last row of  Table~\ref{lo_J}. In fact, comparing the results in the last rows of Tables~\ref{lo_J} and ~\ref{sld_J}, we find that while estimating the field strength, $h$, in presence of interactions, for certain choice of parameters, the LO measurement given in Scheme I can achieve the best metrological precision obtainable using the optimal measurement scheme.

\subsection{Can quantum metrological advantage be restored in absence of interactions?}

\begin{table*}[!htb]
\centering
\begin{tabular}{|c|c|c|c|c|c|}
\hline
   Hamiltonian & \phantom{ami bari}Product\phantom{ami bari} & Maximally entangled & \phantom{ami bari}Product\phantom{ami bari} & Maximally entangled\\ (Estimated parameters) &  (noiseless) &  (noiseless) &  (noisy) &  (noisy) \\ 
  \hline
  \hline  
  $H_4$ ($\omega$) & 0.35 & 0.25 & 0.82 & 0.80 \\
  \hline 
  $H_4$ ($h$) & 0.82 & 0.58 & 0.82 & 0.58 \\ 
  \hline
\end{tabular}
\caption{Minimum uncertainty obtained from Fisher information-based lower bound $\tilde{T}\Delta \tilde{\epsilon}_{min}$ for an estimated parameter $\epsilon$ for various Hamiltonians observed by incorporating some interaction or field terms in the ideal Hamiltonian $\mathcal{H}_2$. The measurement corresponds to Scheme II. The numerical values are correct up to the second decimal place. All numbers in the table are for dimensionless quantities.}
\label{sld_h}
\end{table*}

Here we consider encoding Hamiltonian containing transverse field term in addition to the ideal Hamiltonian.  Therefore, the total Hamiltonian of the system is
\begin{equation}
    H_4=\mathcal{H}_2+\hbar h(\sigma_x^1 + \sigma_x^2).
\end{equation}
The values of the quantity, $\tilde{T}\Delta \tilde{\omega}$, is presented in the  first row of Table~\ref{sld_h} in the noiseless and noisy scenarios. The effect of the transverse field alone is quantitatively the same as that of the transverse field  and interaction considered together in the estimation of $\omega$, i.e., there are instances where optimal maximally entangled probes are better than optimal product ones
both in the noiseless and noisy situations.

Next we consider the quantity, $\tilde{T}\Delta \tilde{h}$, which is given in the last row of Table~\ref{sld_h}. The precision in estimating the field strength, $h$, in absence of interactions, is better if we initiate with an optimal maximally entangled probe instead of optimal product probe - inflicted or not - by noisy environment. Moreover, it is evident that noise has quantitatively no effect in estimating the precision of the field strength, as the quantity, $\tilde{T}\Delta \tilde{h}$, attains the same value both in presence and absence of noise, separately for optimal maximally entangled state and optimal product input state.

\section{Dependence of $\tilde{T}\Delta\tilde{\epsilon}_{min}$ on  entanglement content of  initial states}
\label{arb_in}

\begin{figure}
\centering
\includegraphics[width=8cm]{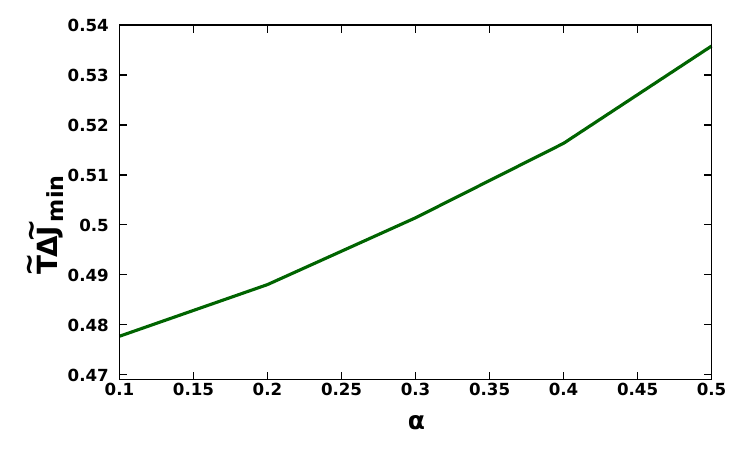}%
\caption{Dependence of $\tilde{T}\Delta \tilde{\epsilon}_{\text{min}}$ on the entanglement content of the initial state. Here we depict $\tilde{T}\Delta\tilde{\omega}_{min}$ vs. $\alpha$ for the noisy case of the system described by the Hamiltonian $H_1$. The measurement is according to Scheme I. All quantities plotted along the horizontal and vertical axes are dimensionless.}
\label{alpha_J}
\end{figure}


In the preceding sections, we have scrutinized the occurrence of metrological advantages in some particular situations with respect to two extreme choices of the initial probes: one was uncorrelated and the other was maximally entangled. As the initial maximal entanglement helps in attaining better measurement precision for some parameters over the initial uncorrelated inputs, while estimating frequency, one can expect that the betterment of precision of estimation of parameters is not a discrete jump from the uncorrelated to maximally entangled initial probes. Instead, a continuous relationship between measurement precision and the entanglement content of the initial probes is anticipated. Hence, how the measurement precision changes with respect to the entanglement content of the initial state may reveal some interesting features. So, we now investigate the behavior of $\Delta \tilde{\epsilon}_{\text{min}}$ for $\epsilon$ being $\omega$, $h$ and $J$, with the increasing entanglement content of the initial states. 
For this purpose, the input state is chosen to be
\begin{equation}
\label{partial_eq}
    \ket{\psi_p^{\prime}}=U_A \otimes U_B \Big[ \alpha \ket{00}+\sqrt{1-\alpha^2}\ket{11} \Big].
\end{equation}
Compare with Eq.~(\ref{initial}). The entanglement content of the state is encapsulated in the parameter $\alpha$. The state is a product one if $\alpha=0$ or $1$, for $\alpha=\frac{1}{\sqrt{2}}$ the state is maximally entangled and between $0$ ($\frac{1}{\sqrt{2}}$) to $\frac{1}{\sqrt{2}}$ $(1)$, it lies in the range between the product (maximally entangled) and maximally entangled (product) states, i.e., a partially entangled one. The local unitaries $U_A$ and $U_B$ will transform one partially entangled state to another without disturbing the entanglement content of the state. In this manner, we get to scan the entire space of partially entangled states for the best choice of the input by optimizing the free parameters of $U_A$ and $U_B$. 
Following the prescription elaborated in the previous sections, we obtain $\tilde{T}\Delta \tilde{\epsilon}_{\text{min}}$ for an estimated parameter $\epsilon$ for different values of $\alpha$ in presence of the dephasing noise considered before. We pertain to measurement scheme I for the derivation of the results in this section.

In contrast to our expectation, we have found that if we concentrate on the estimation of $\omega$ for the Hamiltonian, $H_1$, then, there is no monotonic trend in the curve of $\tilde{T}\Delta \tilde{\omega}_{min}$ vs. $\alpha$. Instead, in this case, we find that the value of $\tilde{T}\Delta \tilde{\omega}_{min}$ is almost constant at $0.82$ for $\alpha=0.1$ to $0.5$ in steps of $0.1$.

Fig.~\ref{alpha_J} depicts the behavior of the minimum uncertainty in the estimation of coupling constant, for the Hamiltonian $H_1$. The optimization has been performed over the input probe states, measurement strategies within Scheme I, and evolution time.
From this figure, it is visible that $\tilde{T}\Delta \tilde{J}_{min}$ increases monotonically from $\alpha=0.1$, until it reaches a minimum for the maximally entangled state corresponding to $\alpha=\frac{1}{\sqrt{2}}$. After that point, the values of $\tilde{T}\Delta \tilde{J}_{\text{min}}$ repeats the nature of the profile and it reaches a maximum again at $\alpha=1$. So, the depiction in this figure explicitly captures how $\tilde{T}\Delta \tilde{J}_{\text{min}}$ varies in the regime in between the product and the maximally entangled states, ultimately attaining a maximum for a maximally entangled one.


\section{Conclusion}
\label{Sec:5}
 It was familiar from previous studies that in absence of noise, one can overcome the shot-noise limit by using a maximally entangled initial state instead of the uncorrelated ones. But this quantum advantage can be lost if we apply a noise to the system, say a local uncorrelated dephasing one. In this paper, we emphasized that the benefit of using maximally entangled probes in quantum metrology, which disappears for frequency estimation in a noisy case, can be restored in presence of two-qubit interactions, like the Ising one, with or without a transverse field, incorporated in the system particles. We note that there were instances, e.g. where frequency estimation in presence of a transverse field was considered and quantum advantage was not restored.



The absence of advantage in using maximally entangled probes was previously reported for estimation of frequency in presence of dephasing noise. We found that the inclusion of field or interaction terms or both can resurrect the
advantage, while still estimating the frequency and while still being acted on by the same dephasing noise. 

We subsequently considered the estimation of the field and interaction strengths, for different system Hamiltonians, and found that while the maximally entangled probe can provide advantage in certain situations, there are also instances where optimal uncorrelated probes prove to be beneficial than maximally entangled ones. 

Finally, we investigated the role of the amount of entanglement of the probe states in quantum parameter estimation. We found that while a monotonically decreasing behavior of the uncertainty of estimation with respect to initial entanglement in the probes is present in some cases,  a constant behavior also crops up in other instances.

 \acknowledgements 
We acknowledge computations performed using Armadillo~\cite{Sanderson,Sanderson1},
NLOPT~\cite{NLOPT}
 (
 ISRES~\cite{Runarsson})
   and QIClib~\cite{QIClib}
   on the cluster computing facility of the Harish-Chandra Research Institute, India. We also acknowledge partial support from the Department of Science and Technology, Government of India through the QuEST grant (grant number DST/ICPS/QUST/Theme-3/2019/120).

\appendix
\section{Joint probabilities of the measurement outcomes for product input probes, and noisy setting}
\label{ap1}
The explicit forms of $p_1$, $p_2$, $p_3$ and $p_4$, defined in Eq.~\eqref{p}, are given by
\begin{widetext}
\begin{eqnarray}
p_1=&& \frac{1}{4}  e^{-i A_1} 
 \Big\{\cos ^2\frac{\theta _1}{2} \cos ^2\frac{\theta _3}{2} e^{i A_2} 
 \Big[4 \cos ^2\frac{\theta _2}{2} \cos ^2\frac{\theta _4}{2} e^{i A_3}+4 \sin ^2\frac{\theta _2}{2} \sin ^2\frac{\theta _4}{2} e^{i A_3}+ \nonumber \\
&& \sin \theta _2 \sin \text{$\theta_4 $} \left(e^{2 i A_4}+e^{2 i \text{$\delta_4 $}}\right)\Big]+ 
 \frac{1}{4} e^{2 i t \text{$\hbar $J}} \Big\{4 \cos ^2\frac{\theta _2}{2} \cos ^2\frac{\theta _4}{2} e^{A_5}
 4 \sin ^2\frac{\theta _1}{2} \sin ^2\frac{\theta _3}{2} e^{i A_6}+\sin \theta _1 \sin \theta _3A_{13}+ \nonumber \\
&& e^{A_{7}} \Big[\sin \theta _1 \sin \theta _3 \Big(4 \sin ^2\frac{\theta _2}{2} \sin ^2\frac{\theta _4}{2} A_8 e^{ A_9}+ 
 \sin \theta _2 \sin \theta _4 A_{10}\Big)+ 
 4  \sin ^2\frac{\theta _1}{2} \sin ^2\frac{\theta _3}{2} \Big(4 \sin ^2\frac{\theta _2}{2} \sin ^2\frac{\theta _4}{2}A_{11}+ \nonumber \\
&& \sin \theta _2 \sin \theta _4 A_{12}\Big)\Big]\Big\}\Big\}, 
\end{eqnarray}
\begin{eqnarray}
p_2=&& \frac{1}{4} e^{-i A_1} \Big\{\cos ^2\frac{\theta _1}{2} \cos ^2\frac{\theta _3}{2} \left(-e^{i A_2}\right) 
 \Big(-4 \sin ^2\frac{\theta _2}{2} \cos ^2\frac{\theta _4}{2} e^{i A_3}-4 \sin ^2\frac{\theta _4}{2} \cos ^2\frac{\theta _2}{2} e^{iA_3}+ \nonumber \\
&& \sin \theta _2 \sin \theta _4 \left(e^{2 i \delta _4}+e^{2 i A_4}\Big)\right)+\frac{1}{4} e^{2 i t \text{$\hbar $J}}  \Big \{ 4 \sin ^2\frac{\theta _2}{2} \cos ^2\frac{\theta _4}{2} e^{A_5}
 \left(4 \sin ^2\frac{\theta _1}{2} \sin ^2\frac{\theta _3}{2} e^{iA_6}+\sin \theta _1 \sin \theta _3 A_8 \right)+e^{A_{7}} \nonumber \\
&&\Big[4 \cos ^2\frac{\theta _2}{2} \Big(4 \sin ^2\frac{\theta _1}{2}\sin ^2\frac{\theta _3}{2} A_{11}   \sin \theta _1 \sin \theta _3 + e^{A_9}A_{13}\Big) 
 \sin ^2\frac{\theta _4}{2}-\sin \theta _2 \left(4 \sin ^2\frac{\theta _1}{2} \sin ^2\frac{\theta _3}{2} A_{12} +\right. \nonumber \\
&& \sin \theta _1 \sin \theta _3\ A_{10} \Big) \sin \theta _4\Big]\Big \}\Big\}, 
\end{eqnarray}
\begin{eqnarray}
p_3=&& \frac{1}{4} e^{-i A_1} \Big\{\cos ^2\frac{\theta _2}{2} \cos ^2\frac{\theta _4}{2} \left(-e^{i A_{14}}\right) 
 \Big(-4 \sin ^2\frac{\theta _1}{2} \cos ^2\frac{\theta _3}{2} e^{i A_{15}}-4 \sin ^2\frac{\theta _3}{2} \cos ^2\frac{\theta _1}{2} e^{i A_{15}}+ \nonumber \\
&& \sin \theta _1 \sin \theta _3 A_{13}\Big)+\frac{1}{4} e^{2 i t \text{$\hbar $J}} \Big\{4 \sin ^2\frac{\theta _1}{2} \cos ^2\frac{\theta _3}{2} A_{17} 
 4 \sin ^2\frac{\theta _2}{2} \sin ^2\frac{\theta _4}{2} e^{i A_3}+\sin \theta _2 \sin \theta _4 C_1+ \nonumber \\
&& e^{3 \gamma  t+2 i t \hbar \omega } \Big[-\sin \theta _1 \sin \theta _3 \Big(4 \sin ^2\frac{\theta _2}{2} \sin ^2\frac{\theta _4}{2}
e^{A_9} A_8+
 \sin \theta _2 \sin \theta _4 A_{10} \Big)+ \nonumber \\
&& 4 \cos ^2\frac{\theta _1}{2} \sin ^2\frac{\theta _3}{2} \Big(4 \sin ^2\frac{\theta _2}{2} \sin ^2\frac{\theta _4}{2} A_{11}+
 \sin \theta _2 \sin \theta _4 e^{i \left(\delta _1+\delta _3\right)} A_{16} \Big)\Big]\Big\}\Big\}, 
\end{eqnarray}
\begin{eqnarray}
p_4=&& \frac{1}{4} e^{-i A_1} \Big\{\sin ^2\frac{\theta _1}{2} \cos ^2\frac{\theta _3}{2} \left(-e^{i A_2}\right) 
 \left(-4 \sin ^2\frac{\theta _2}{2} \cos ^2\frac{\theta _4}{2} e^{i A_3}-4 \sin ^2\frac{\theta _4}{2} \cos ^2\frac{\theta _2}{2} e^{i A_3}+\right. \nonumber \\
&& \sin \theta _2 \sin \theta _4 C_1\Big)+\frac{1}{4} e^{t2i\text{$\hbar $J}}e^{A_7} 
 \Big\{-\sin \theta _1 \sin \theta _3 \Big(4 \sin ^2\frac{\theta _2}{2} \cos ^2\frac{\theta _4}{2} e^{A_9} A_8 + \nonumber \\
&& 4 \sin ^2\frac{\theta _4}{2} \cos ^2\frac{\theta _2}{2} e^{A_9} A_{13}- 
 \sin \theta _2 \sin \theta _4 A_{10}\Big)+ 
 4 \cos ^2\frac{\theta _1}{2} \sin ^2\frac{\theta _3}{2} \Big[4 \sin ^2\frac{\theta _2}{2} \cos ^2\frac{\theta _4}{2} A_{11}+ \nonumber \\
&&  e^{i \left(\delta _1+\delta _3\right)}  \Big(4 \sin ^2\frac{\theta _4}{2} \cos ^2\frac{\theta _2}{2} e^{i A_3}e^{t (\gamma +i \hbar \omega )}-\sin \theta _2 \sin \theta _4 A_{16} \Big)\Big]\Big\}\Big\},
\end{eqnarray}
\end{widetext}
where \(A_1=\left(\delta _1+\delta _2+\delta _3+\delta _4-5 i \gamma  t+4 t \text{$\hbar $J}+4 t \hbar \omega \right)\), 
\(A_2 = \left(\delta _1+\delta _3-4 i \gamma  t+2 t \text{$\hbar $J}+3 t \hbar \omega \right)\), 
\(A_3=\left(\delta _2+\delta _4-i \gamma  t+2 t \text{$\hbar $J}+t \hbar \omega \right)\), \(A_4=\left(\delta _2+2 t \text{$\hbar $J}+t \hbar \omega \right)\), \(A_5=i \left(\delta _2+\delta _4\right)+t (4 \gamma +3 i \hbar \omega )\), \(A_6=\left(\delta _1+\delta _3-i \gamma  t+2 t \text{$\hbar $J}+t \hbar \omega \right)\), \(A_{7}=3 \gamma  t+2 i t \hbar \omega \), \(A_8=\left(e^{2 i \left(\delta _3+2 t \text{$\hbar $J}\right)}+e^{2 i \left(\delta _1+t \hbar \omega \right)}\right)\), \(A_9=i\left(\delta _2+\text{$\delta_4$}\right)+t (\gamma +i \hbar \omega )\), \(A_{10}=e^{2 i t \text{$\hbar $J}} \left(e^{2 i \delta _3}+e^{2 i \left(\delta _1+t \hbar \omega \right)}\right) \left(e^{2 i \left(\delta _2+t \hbar \omega \right)}+e^{2 i \text{$\delta_4 $}}\right)\), \(A_{11}=e^{i \left(\delta _1+\delta _3\right)} e^{i \left(\delta _2-2 i \gamma  t+2 t \text{$\hbar $J}+2 t \hbar \omega +\text{$\delta_4$}\right)}\) \(A_{12}=e^{i \left(\delta _1+\delta _3\right)} e^{t (\gamma +i \hbar \omega )} \left(e^{2 i \left(\delta _4+2 t \text{$\hbar $J}\right)}+e^{2 i \left(\delta _2+t \hbar \omega \right)}\right)\) \(=e^{i \left(\delta _1+\delta _3\right)} e^{t (\gamma +i \hbar \omega )}C_1\), \(A_{13}= \left(e^{2 i \delta _3}+e^{2 i \left(\delta _1+2 t \text{$\hbar $J}+t \hbar \omega \right)}\right)\), 
\(A_{14}=\left(\delta _2+\delta _4-4 i \gamma  t+2 t \text{$\hbar $J}+3 t \hbar \omega \right)\), \(A_{15}=\left(\delta _1+\delta _3-i \gamma  t+2 t \text{$\hbar $J}+t \hbar \omega \right)\), \(A_{16}=e^{t (\gamma +i \hbar \omega )} \left(e^{2 i \delta _4}+e^{2 i \left(\delta _2+2 t \text{$\hbar $J}+t \hbar \omega \right)}\right)\) and \(A_{17}=e^{i \left(\delta _1+\delta _3\right)+t (4 \gamma +3 i \hbar \omega )}\).

\end{document}